\documentclass{elsarticle}
\usepackage{amsmath}
\usepackage{natbib}
\journal{International journal of Non-Linear Mechanics}

\begin{document}

\begin{frontmatter}

\title{Conservation Laws of the Two-Dimensional Gas Dynamics Equations}

\author{E.I.~Kaptsov$^{1,2}$}

\author[rvt]{S.V.~Meleshko \corref{cor1}}

\cortext[cor1]{Corresponding author}

\address{$^{1}$School of Mathematics, Institute of Science, Suranaree University
of Technology, Nakhon Ratchasima, 30000, Thailand}

\ead{sergey@math.sut.ac.th}

\address{$^{2}$Keldysh Institute of Applied Mathematics, \\
 Russian Academy of Science, Miusskaya Pl. 4, Moscow, 125047, Russia
\\
 }
\begin{abstract}
Two-dimensional gas dynamics equations in mass Lagrangian coordinates
are studied in this paper. The equations describing these flows are reduced
to two Euler-Lagrange equations. Using group classification and Noether's
theorem, conservation laws are obtained. Their counterparts in Eulerian
coordinates are given. Among these counterparts there are new conservation laws.
\end{abstract}
\end{frontmatter}

%\pagebreak{}

%\section{Corrections}

%1. $X^e_8$

%2. Added Clarkson

%3. In Lagrangian added 1 in degree.

\section{Introduction}

\subsection{Lagrangian fluid dynamics}

Physical phenomena in continuum mechanics are modeled in two distinct ways. The typical approach for fluid dynamics uses
Eulerian coordinates, where the system describes fluid motion at fixed
locations. Velocities, density and other properties of fluid particles
in the Eulerian description are considered to be functions of time
and of fixed space coordinates. In contrast, in the Lagrangian description,
particles are identified by the positions which they
occupy at some initial time. Typically, Lagrangian coordinates are
not applied in the description of fluid motion. However, in some special
contexts the Lagrangian description is indeed useful in solving certain
problems. Analytically, both coordinate systems are capable of producing
exact solutions of fluid flows, including discontinuous flows. They
are regarded as equivalent to another, except that the Lagrangian
system gives more information: `in practice such description is often
too detailed and complicated, but it is always implied in formulating
physical laws' \citep{bk:Sedov[mss]}.

There is extensive literature on Lagrangian fluid dynamics
(see, for example, \citep{bk:Webb2018,bk:WebbZank[2009],bk:DespresMazeran2005}
and references therein). These studies can be roughly separated out
into two groups: the analysis of equations in the gas dynamics variables in mass Lagrangian
coordinates \citep{bk:RozhdYanenko[1978],bk:DespresMazeran2005} and
analysis of Euler-Lagrange equations obtained by varying the Lagrangian
in mass Lagrangian coordinates \citep{bk:WebbZank[2009],bk:Webb2018,bk:SiriwatKaewmaneeMeleshko2016,%
bk:VorakaKaewmaneeMeleshko2019}.
In the present paper the Euler-Lagrange equations in mass Lagrangian
coordinates of the gas dynamics equations are studied.

\subsection{Symmetries and conservation laws}

Symmetries have always attracted the attention of scientists. One
of the tools for studying symmetries is the group analysis method
\citep{bk:Ovsiannikov1978,bk:Olver[1986],bk:MarsdenRatiu[1994],bk:Ibragimov[1999],bk:Cantwell[2002]},
which is a basic method for constructing exact solutions of partial
differential equations. The group properties of the gas dynamics equations
in Eulerian coordinates were studied in \citep{bk:Ovsiannikov1978,bk:Ovsiannikov[1994]}.
Extensive group analysis of the one-dimensional gas dynamics equations
in mass Lagrangian coordinates was given in \citep{bk:AkhatovGazizovIbragimov[1991],bk:HandbookLie_v2}.
Here the results of \citep{bk:SjobergMahomed2004} should also be mentioned,
where nonlocal conservation laws of the one-dimensional gas dynamics
equations in mass Lagrangian coordinates were found. The authors of
\citep{bk:WebbZank[2009],bk:Webb2018} analyzed the Euler-Lagrange
equations corresponding to the one-dimensional gas dynamics equations
in mass Lagrangian coordinates and extensions of the known conservation
laws were derived. These conservation laws correspond to special forms
of the entropy. The group nature of these conservation laws is given
in \citep{bk:DorodnitsynKozlovMeleshko2019}, where a complete group
classification of these Euler-Lagrange equations is presented. Notice that
the hyperbolic shallow water equations are equivalent to the isentropic
gas dynamics equations of a polytropic gas with $\gamma=2$. The complete
group analysis of the Euler-Lagrange equations of the one-dimensional
gas dynamics equations of isentropic flows of a polytropic gas with $\gamma=2$ was given in \citep{bk:SiriwatKaewmaneeMeleshko2016}. Group properties of the two-dimensional shallow water equations describing flows over a bed which is rotating
with position-dependent Coriolis parameter in Lagrangian coordinates are studied in \citep{bk:BilaMansfieldClarkson2006}. Conservation laws were constructed by using a Lagrangian of the form presented in \citep{bk:Salmon1983}.

As mentioned above, besides assisting with the construction of exact
solutions, the knowledge of an admitted Lie group allows one to derive
conservation laws. Conservation laws provide information on the basic
properties of solutions of differential equations, and they are also
needed in the analyses of stability and global behavior of solutions.
Noether's theorem \citep{bk:Noether[1918]} is the tool which relates
symmetries and conservation laws. However, the application of Noether's
theorem depends on the following condition: that the differential equations
under consideration can be rewritten as Euler-Lagrange equations with
appropriate Lagrangian. Among approaches which try to overcome this
limitation one can mention here the approaches developed in \citep{bk:Shmyglevski,Ibragimov2[2007],bk:BlumanCheviakovAnco,%
bk:SeligerWhitham[1968]}\footnote{Therein one can find more details and references.}.

\subsection{Noether's theorem}

We begin with the background related to the application of symmetries
for constructing conservation laws\footnote{The reader is referred to \citep{bk:Ibragimov[2010]} for the details
on symmetries and conservation laws.}.

Let
\[
X=\xi^{i}\frac{\partial}{\partial x^{i}}+\eta^{k}\frac{\partial}{\partial u^{k}}+\eta_{i}^{k}\frac{\partial}{\partial u_{i}^{k}}+...
\]
be a Lie-B\"acklund operator, and $F=F(x,u,p)$. Here $x=(x^{1},x^{2},\ldots,x^{n})$
are the independent variables, $u=(u^{1},u^{2},\ldots,u^{m})$ are
the dependent variables, and $p$ denotes their derivatives $u_{i_{1}...i_{l}}^{k}$
of finite order.

Noether's theorem is based on two identities. The first identity is
called the Noether identity \citep{bk:Ibragimov[2010]},
\begin{equation}
XF+FD_{i}\xi^{i}=W^{k}\frac{\delta F}{\delta u^{k}}+D_{i}({\cal N}^{i}F)
\label{eq:Noether}
\end{equation}
where
\begin{equation}
\frac{\delta}{\delta u^{k}}=\frac{\partial}{\partial u^{k}}+\sum_{s=1}(-1)^{s}D_{i_{1}}...D_{i_{s}}\frac{\partial}{\partial u_{i_{1}i_{2}...i_{s}}^{k}},\,\,\,(k=1,2,...,m),\label{eq:variational_1}
\end{equation}
are variational derivatives,
\[
W^{k}=\eta^{k}-\xi^{i}u_{i}^{k},\,\,\,(k=1,2,...,m),
\]
and
\[
{\cal N}^{i}F=\xi^{i}F+W^{k}\frac{\delta F}{\delta u_{i}^{k}}+\sum_{s=1}D_{i_{1}}...D_{i_{s}}(W^{k})\frac{\delta F}{\delta u_{ii_{1}i_{2}...i_{s}}^{k}},\,\,\,(i=1,2,...,n).
\]
Here the variational derivatives ${\displaystyle \frac{\delta F}{\delta u_{i_{1}i_{2}...i_{s}}^{k}}}$
are obtained from (\ref{eq:variational_1}) by replacing $u^{k}$
with the corresponding derivative $u_{i_{1}i_{2}...i_{s}}^{k}$. The
second identity is \citep{bk:Olver[1986],bk:DorodnitsynKozlov[2011]}\footnote{For the sake of simplicity this identity is presented in the case,
where a function $F(x,u,p)$ does not depend on derivatives of order
2 and higher, and the coefficients do not depend on derivatives $\xi^{i}=\xi^{i}(x,u)$,
$\eta^{k}=\eta^{k}(x,u)$. Identity (\ref{eq:second}) is valid in
more general cases. }: %\footnote{In \citep{bk:DorodnitsynKozlov[2011]}, identity
%(\ref{eq:second})is given with $n=1$.}:
\begin{equation}
\begin{array}{c}
{\displaystyle \frac{\delta}{\delta u^{j}}\left(XF+FD_{i}\xi^{i}-D_{i}B^{i}\right)=X(\frac{\delta F}{\delta u^{j}})+\frac{\delta F}{\delta u^{k}}\left(\frac{\partial\eta^{k}}{\partial u^{j}}-\frac{\partial\xi^{i}}{\partial u^{j}}u_{i}^{k}+\delta_{kj}D_{i}\xi^{i}\right),}\\
(j=1,2,...,m).
\end{array}\label{eq:second}
\end{equation}

\textbf{Theorem} (Noether). Suppose, a generator
\[
X=\xi^{i}(x,u)\frac{\partial}{\partial x^{i}}+\eta^{k}(x,u)\frac{\partial}{\partial u^{k}}
\]
satisfies the equation
\begin{equation}
X{\cal L}+{\cal L}D_{i}\xi^{i}=D_{i}B^{i}.\label{eq:divergent}
\end{equation}
Then the generator $X$ is an admitted symmetry of the system of
Euler-Lagrange equations
\begin{equation}
\frac{\delta{\cal L}}{\delta u^{k}}=0,\,\,\,(k=1,2,...,m),\label{eq:Euler-Lagrange-1}
\end{equation}
and the vector
\[
({\cal N}^{1}{\cal L}-B^{1},{\cal N}^{2}{\cal L}-B^{2},...{\cal N}^{n}{\cal L}-B^{n})
\]
is a conserved vector.

In the case $\boldsymbol{B}=(B^{1},B^{2},...,B^{n})=0$, the symmetry
$X$ is called a variational symmetry, otherwise for $\boldsymbol{B}\neq0$
the symmetry $X$ is called a divergent one.

\subsection{Objectives of the present paper}

The present paper is focused on the two-dimensional Euler-Lagrange
gas dynamics equations of a polytropic gas. Its objective
is to make group classification of the Euler-Lagrange equations with
respect to the function of entropy, and to construct conservation
laws by applying Noether's theorem.

\section{Gas dynamics equations of a polytropic gas}

In this section the gas dynamics equations of a polytropic gas in
mass Lagrangian coordinates are considered. The two-dimensional Euler-Lagrange
equations are obtained by using a variational approach.

\subsection{Gas dynamics equations in Eulerian coordinates}

The gas dynamics equations of a polytropic gas are
\begin{equation}
\rho_{t}+u\rho_{x}+v\rho_{y}+\rho(u_{x}+v_{y})=0,\label{eq:cons_mass_E}
\end{equation}
\begin{equation}
\rho(u_{t}+uu_{x}+vu_{y})+p_{x}=0,\label{eq:lin_mom_1}
\end{equation}
\begin{equation}
\rho(v_{t}+uv_{x}+vv_{y})+p_{y}=0,\label{eq:lin_mom_2}
\end{equation}
\begin{equation}
S_{t}+uS_{x}+vS_{y}=0,\label{eq:entropy_E}
\end{equation}
where $\rho$ is the density, $(u,v)$ is the velocity,
\[
p=S\rho^{\gamma},
\]
and $S$ is a function depending on the entropy $\tilde{S}$ of
the polytropic gas. The function $S$ is related with the entropy $\tilde{S}$
of a polytropic gas as follows \citep{bk:Ovsiannikov[2003]}
\[
S=Re^{(\tilde{S}-\tilde{S}_{0})/c_{v}},
\]
where $R$ is the gas constant, $c_{v}$ is the dimensionless specific
heat capacity at constant volume, and $\tilde{S}_{0}$ is constant.

\subsection{Lagrangian coordinates}

The relations
\[
x=\varphi_{1}(t,\tilde{\xi},\tilde{\eta}),\,\,\,y=\varphi_{2}(t,\tilde{\xi},\tilde{\eta})
\]
between the Lagrangian $(t,\tilde{\xi},\tilde{\eta})$ and Eulerian
$(t,x,y)$ coordinates are defined by the equations
\[
\varphi_{1t}(t,\tilde{\xi},\tilde{\eta})=u(t,\varphi_{1}(t,\tilde{\xi},\tilde{\eta}),\varphi_{2}(t,\tilde{\xi},\tilde{\eta})),\,\,\,\varphi_{2t}(t,\tilde{\xi},\tilde{\eta})=v(t,\varphi_{1}(t,\tilde{\xi},\tilde{\eta}),\varphi_{2}(t,\tilde{\xi},\tilde{\eta})).
\]
In Lagrangian coordinates the entropy is an arbitrary function of
$\tilde{\xi}$ and $\tilde{\eta}$. The conservation law of mass provides
the relation \citep{bk:Sedov[mss]}:
\[
\rho=\frac{\rho_{0}}{\varphi_{1\tilde{\xi}}\varphi_{2\tilde{\eta}}-\varphi_{1\tilde{\eta}}\varphi_{2\tilde{\xi}}},
\]
where $\rho_{0}=\rho_{0}(\tilde{\xi},\tilde{\eta})>0$ is the function
of integration.

Applying the change
\[
\xi=h_{1}(\tilde{\xi},\tilde{\eta}),\,\,\,\eta=h_{2}(\tilde{\xi},\tilde{\eta}),
\]
one finds that
\[
\varphi_{1\tilde{\xi}}\varphi_{2\tilde{\eta}}-\varphi_{1\tilde{\eta}}\varphi_{2\tilde{\xi}}=(\varphi_{1\xi}\varphi_{2\eta}-\varphi_{1\eta}\varphi_{2\xi})(h_{1\tilde{\xi}}h_{2\tilde{\eta}}-h_{1\tilde{\eta}}h_{2\tilde{\xi}}).
\]
Hence, choosing $h_{1}(\tilde{\xi},\tilde{\eta})$ and $h_{2}(\tilde{\xi},\tilde{\eta})$
such that
\[
h_{1\tilde{\xi}}h_{2\tilde{\eta}}-h_{1\tilde{\eta}}h_{2\tilde{\xi}}=\rho_{0},
\]
one derives that
\[
\rho(t,\xi,\eta)=J^{-1}(t,\xi,\eta),
\]
where $J=\varphi_{1\xi}\varphi_{2\eta}-\varphi_{1\eta}\varphi_{2\xi}\neq0$.
Following the one-dimensional case, the coordinates $(t,\xi,\eta)$
are called the mass Lagrangian coordinates. The gas dynamics equations
(\ref{eq:cons_mass_E})-(\ref{eq:entropy_E}) in the mass Lagrangian
coordinates with the gas dynamics variables become \citep{bk:DespresMazeran2005}:
\begin{equation}
\left(\frac{1}{\rho}\right)_{t}=\tilde{u}_{\xi}\varphi_{2\eta}
-\varphi_{1\eta}\tilde{v}_{\xi}-(\tilde{u}_{\eta}\varphi_{2\xi}
-\varphi_{1\xi}\tilde{v}_{\eta}),
\label{eq:cons_mass_L}
\end{equation}
\begin{equation}
\tilde{u}_{t}+\varphi_{2\eta}\tilde{p}_{\xi}-\varphi_{2\xi}\tilde{p}_{\eta}=0,
\label{eq:lin_mom_1L}
\end{equation}
\begin{equation}
\tilde{v}_{t}-\varphi_{1\eta}\tilde{p}_{\xi}+\varphi_{1\xi}\tilde{p}_{\eta}=0,
\label{eq:lin_mom_2L}
\end{equation}
\begin{equation}
\tilde{S}_{t}=0,
\label{eq:entropy_L}
\end{equation}
\begin{equation}
\varphi_{1t}=\tilde{u},\,\,\,\varphi_{2t}=\tilde{v}.
\label{eq:geometrical}
\end{equation}
The gas dynamics variable $\tilde{f}(t,\xi,\eta)$ in mass Lagrangian
coordinates and in Eulerian coordinates $f(t,x,y)$ are related by
the formulae
\[
\tilde{f}(t,\xi,\eta)=f(t,\varphi_{1}(t,\tilde{\xi},\tilde{\eta}),\varphi_{2}(t,\tilde{\xi},\tilde{\eta})).
\]
As there is no ambiguity, the sign $\tilde{}$ is further omitted. Notice
that in the one-dimensional case one assumes that
\[
\varphi_{1}=\varphi_{1}(t,\xi),\,\,\,\varphi_{2}=\eta,
\]
the system of equations (\ref{eq:cons_mass_L})-(\ref{eq:entropy_L})
is closed, and it becomes the classical one-dimensional gas dynamics
equations in mass Lagrangian coordinates \citep{bk:RozhdYanenko[1978]}:
\begin{equation}
\left(\frac{1}{\rho}\right)_{t}=u_{\xi},\ \ \ u_{t}+p_{\xi}=0,\ \ \ S_{t}=0.\label{eq:cons_L-1d}
\end{equation}

\subsection{Variational approach}

Choosing the Lagrangian
\[
{\cal L}=\frac{\varphi_{1t}^{2}+\varphi_{2t}^{2}}{2}-\frac{1}{\gamma-1}J^{1-\gamma}S,
\]
one derives the Euler-Lagrange equations ${\displaystyle \frac{\delta L}{\delta\varphi_{1}}}=0$
and ${\displaystyle \frac{\delta L}{\delta\varphi_{2}}}=0$:
\begin{equation}
\begin{array}{c}
J^{\gamma}\varphi_{1tt}+S_{\xi}\varphi_{2\eta}-S_{\eta}\varphi_{2\xi}+\gamma J^{-1}S\left(\varphi_{2\eta}(\varphi_{1\eta}\varphi_{2\xi\xi}-\varphi_{2\eta}\varphi_{1\xi\xi})\right.\\
\left.+\varphi_{2\xi}(\varphi_{1\xi}\varphi_{2\eta\eta}-\varphi_{1\eta\eta}\varphi_{2\xi})+2\varphi_{2\xi}\varphi_{2\eta}\varphi_{1\xi\eta}-(\varphi_{1\xi}\varphi_{2\eta}+\varphi_{1\eta}\varphi_{2\xi})\varphi_{2\xi\eta}\right)=0,
\end{array}\label{eq:gas_L_1}
\end{equation}
\begin{equation}
\begin{array}{c}
J^{\gamma}\varphi_{2tt}-S_{\xi}\varphi_{1\eta}+S_{\eta}\varphi_{1\xi}+\gamma SJ^{-1}\left(\varphi_{1\eta}(\varphi_{2\eta}\varphi_{1\xi\xi}-\varphi_{1\eta}\varphi_{2\xi\xi})\right.\\
\left.+\varphi_{1\xi}(\varphi_{2\xi}\varphi_{1\eta\eta}-\varphi_{1\xi}\varphi_{2\eta\eta})+2\varphi_{1\xi}\varphi_{1\eta}\varphi_{2\xi\eta}-(\varphi_{1\xi}\varphi_{2\eta}+\varphi_{1\eta}\varphi_{2\xi})\varphi_{1\xi\eta}\right)=0.
\end{array}\label{eq:gas_L_2}
\end{equation}

One can show that equations (\ref{eq:gas_L_1}), (\ref{eq:gas_L_2})
are reduced to the equations of linear momentum (\ref{eq:lin_mom_1})
and (\ref{eq:lin_mom_2}). Corresponding changes of derivatives are
obtained from the identities
\[
\varphi_{1t}(t,\xi,\eta)=u(t,\varphi_{1}(t,\xi,\eta),\varphi_{2}(t,\xi,\eta)),\,\,\,
\varphi_{2t}(t,\xi,\eta)=v(t,\varphi_{1}(t,\xi,\eta),\varphi_{2}(t,\xi,\eta)),
\]
\[
J(t,\xi,\eta)=\rho^{-1}(t,\varphi_{1}(t,\xi,\eta),\varphi_{2}(t,\xi,\eta)).
\]
For example,
\[
\varphi_{1t\xi}=\varphi_{1\xi}u_{x}+\varphi_{2\xi}u_{y},\,\,\,\,\,\,\varphi_{2t\xi}=\varphi_{1\xi}v_{x}+\varphi_{2\xi}v_{y},
\]
\[
\varphi_{1t\eta}=\varphi_{1\eta}u_{x}+\varphi_{2\eta}u_{y},\,\,\,\,\,\,\varphi_{2t\eta}=\varphi_{1\eta}v_{x}+\varphi_{2\eta}v_{y},
\]
\[
\varphi_{1tt}=u_{t}+uu_{x}+vu_{y},\,\,\,\,\,\,\varphi_{2tt}=v_{t}+uv_{x}+vv_{y},
\]
\[
S_{\xi}=\varphi_{1\xi}S_{x}+\varphi_{2\xi}S_{y},\,\,\,S_{\eta}=\varphi_{1\eta}S_{x}+\varphi_{2\eta}S_{y}.
\]
\[
\varphi_{1tt}=u_{t}+uu_{x}+vu_{y},\,\,\,\,\,\,\varphi_{2tt}=v_{t}+uv_{x}+vv_{y},
\]
\[
J_{\xi}=-\rho^{-2}(\varphi_{1\xi}\rho_{x}+\varphi_{2\xi}\rho_{y}),\,\,\,J_{\eta}=-\rho^{-2}(\varphi_{1\eta}\rho_{x}+\varphi_{2\eta}\rho_{y}),
\]
\[
J_{t}=-\rho^{-2}(\rho_{t}+u\rho_{x}+v\rho_{y}).
\]

In the present paper equations (\ref{eq:gas_L_1}), (\ref{eq:gas_L_2})
are called the gas dynamics equations in mass Lagrangian coordinates,
whereas equations (\ref{eq:cons_mass_L})-(\ref{eq:entropy_L}) are
called the gas dynamics equations in the gas dynamics variables in
mass Lagrangian coordinates.

\subsection{Relations between conserved vectors}

As there is the change

\[
D_{\xi}T=\varphi_{1\xi}D_{x}T+\varphi_{2\xi}D_{y}T,\,\,\,D_{\eta}T=\varphi_{1\eta}D_{x}T+\varphi_{2\eta}D_{y}T,
\]
one has
\[
D_{\xi}T=J\left(D_{x}\left(\frac{\varphi_{1\xi}}{J}T\right)+D_{y}\left(\frac{\varphi_{2\xi}}{J}T\right)\right),\,\,\,D_{\eta}T=J\left(D_{x}\left(\frac{\varphi_{1\eta}}{J}T\right)+D_{y}\left(\frac{\varphi_{2\eta}}{J}T\right)\right).
\]
Because of the following relations
\[
\begin{array}{c}
D_{t}T^{t}+D_{\xi}T^{\xi}+D_{\eta}T^{\eta}=D_{t}(\rho JT^{t})+D_{\xi}T^{\xi}+D_{\eta}T^{\eta}
\\[2ex]
=\rho T^{t}\frac{dJ}{dt}+JD_{t}(\rho T^{t})+D_{\xi}T^{\xi}+D_{\eta}T^{\eta}
\\[2ex]
=\rho\frac{dJ}{dt}T^{t}+J\left(D_{\bar{t}}(\rho T^{t})+uD_{x}(\rho T^{t})+vD_{y}(\rho T^{t})\right)
\\
+\varphi_{1\xi}D_{x}T^{\xi}+\varphi_{2\xi}D_{y}T^{\xi}+\varphi_{1\eta}D_{x}T^{\eta}
+\varphi_{2\eta}D_{y}T^{\eta}
\\[2ex]
=\rho T^{t}\frac{dJ}{dt}+J\left(D_{\bar{t}}(\rho T^{t})+(D_{x}(\rho uT^{t})-u_{x}\rho T^{t})+(D_{y}(\rho vT^{t})-v_{y}\rho T^{t})\right)
\\
+\varphi_{1\xi}D_{x}T^{\xi}+\varphi_{2\xi}D_{y}T^{\xi}
+\varphi_{1\eta}D_{x}T^{\eta}+\varphi_{2\eta}D_{y}T^{\eta}
\\[2ex]
\left.
=\rho T^{t}\left(\frac{dJ}{dt}-J\,(u_{x}+v_{y})\right)+J\left(D_{\bar{t}}(\rho T^{t})+D_{x}(\rho uT^{t})+D_{y}(\rho vT^{t})
\right.
\right)
\\
+J\left(D_{x}\left(\frac{\varphi_{1\xi}}{J}T^{\xi}\right)
+D_{y}\left(\frac{\varphi_{2\xi}}{J}T^{\xi}\right)
+D_{x}\left(\frac{\varphi_{1\eta}}{J}T^{\eta}\right)
+D_{y}\left(\frac{\varphi_{2\eta}}{J}T^{\eta}\right)\right)
\\[2ex]
=J\left(D_{\bar{t}}(\rho T^{t})+D_{x}(\rho uT^{t})+D_{y}(\rho vT^{t})
\right.
\\
\left.
+D_{x}\left(\frac{\varphi_{1\xi}}{J}T^{\xi}\right)
+D_{y}\left(\frac{\varphi_{2\xi}}{J}T^{\xi}\right)
+D_{x}\left(\frac{\varphi_{1\eta}}{J}T^{\eta}\right)
+D_{y}\left(\frac{\varphi_{2\eta}}{J}T^{\eta}\right)\right)
\\[2ex]
=J\left(D_{\bar{t}}(\rho T^{t})+D_{x}\left(\rho uT^{t}+\frac{\varphi_{1\xi}}{J}T^{\xi}+\frac{\varphi_{1\eta}}{J}T^{\eta}\right)+D_{y}\left(\rho vT^{t}+\frac{\varphi_{2\xi}}{J}T^{\xi}+\frac{\varphi_{2\eta}}{J}T^{\eta}\right)\right)
\end{array}
\]
one finds that a conserved vector $(T^{t},T^{\xi},T^{\eta})$ in Lagrangian
coordinates reduces to the conserved vector $(^{e}T^{t},T^{x},T^{y})$
in Eulerian coordinates, where
\[
^{e}T^{t}=\rho T^{t},\,\,\,T^{x}=\rho uT^{t}+\frac{\varphi_{1\xi}}{J}T^{\xi}+\frac{\varphi_{1\eta}}{J}T^{\eta},\,\,\,T^{y}=\rho vT^{t}+\frac{\varphi_{2\xi}}{J}T^{\xi}+\frac{\varphi_{2\eta}}{J}T^{\eta}.
\]

\section{Group classification of equations (\ref{eq:gas_L_1}), (\ref{eq:gas_L_2})}

The group classification consists of two steps \citep{bk:Ovsiannikov1978}.
For the first step in the group classification one needs to find an
equivalence Lie group which can be used for simplification of the arbitrary
elements contained in the equations studied. These simplifications are
used in analysis of classifying equations, a general solution of which
provides admitted Lie groups and representations of simplified arbitrary
elements.

\subsection{Equivalence transformations}

The group classification of equations depends on the representations of the
arbitrary elements \citep{bk:Ovsiannikov1978}. As the function $S(\xi,\eta)$
is an arbitrary element of equations (\ref{eq:gas_L_1}), (\ref{eq:gas_L_2}),
the group classification has to be made with respect to it.

The generator of the equivalence transformations is considered in
the form
\[
X^{e}=\zeta^{t}\frac{\partial}{\partial t}+\zeta^{\xi}\frac{\partial}{\partial\xi}+\zeta^{\eta}\frac{\partial}{\partial\eta}+\zeta^{\varphi_{1}}\frac{\partial}{\partial\varphi_{1}}+\zeta^{\varphi_{2}}\frac{\partial}{\partial\varphi_{2}}+\zeta^{S}\frac{\partial}{\partial S},
\]
where all coefficients depend on $(t,\xi,\eta,\varphi_{1},\varphi_{2},S)$.
Calculations show that a basis of the Lie algebra corresponding to
the equivalence Lie group consists of the generators
\[
X_{1}^{e}=\partial_{\varphi_{1}},\,\,\,X_{2}^{e}=\partial_{\varphi_{2}},\,\,\,X_{3}^{e}=t\partial_{\varphi_{1}},\,\,\,X_{4}^{e}=t\partial_{\varphi_{2}},\,\,\,X_{5}^{e}=\varphi_{2}\partial_{\varphi_{1}}-\varphi_{1}\partial_{\varphi_{2}},\,\,\,X_{6}^{e}=\partial_{t},
\]
\[
X_{7}^{e}=\varphi_{1}\partial_{\varphi_{1}}+\varphi_{2}\partial_{\varphi_{2}}+2\gamma S\partial_{S},\,\,\,X_{8}^{e}=\xi\partial_{\xi}+\eta\partial_{\eta}+2(1-\gamma)S\partial_{S},
\]
\[
X_{9}^{e}=t\partial_{t}-2S\partial_{S},\,\,\,X_{\psi}^{e}=-\psi_{\eta}\partial_{\xi}+\psi_{\xi}\partial_{\eta},
\]
where $\psi(\xi,\eta)$ is an arbitrary function. For $\gamma=2$
there is an additional set of equivalence transformations corresponding
to the generator

\[
X_{10}^{e}=t(t\partial_{t}+\varphi_{1}\partial_{\varphi_{1}}+\varphi_{2}\partial_{\varphi_{2}}).
\]

\subsection{Classifying equation}

The admitted Lie group is derived by solving the determining equations
\begin{equation}
\tilde{X}E_{i|E=0}=0,\,\,\,(i=1,2),\label{eq:determining}
\end{equation}
where $E_{i}$ is the left-hand side of equations (\ref{eq:gas_L_1}),
(\ref{eq:gas_L_2}), $|E=0$ means that the expressions $XE_{i}$
are considered on the manifold defined by equations (\ref{eq:gas_L_1}),
(\ref{eq:gas_L_2}), and the generator $\tilde{X}$ is the prolongation
of the generator
\[
X=\zeta^{t}\frac{\partial}{\partial t}+\zeta^{\xi}\frac{\partial}{\partial\xi}+\zeta^{\eta}\frac{\partial}{\partial\eta}+\zeta^{\varphi_{1}}\frac{\partial}{\partial\varphi_{1}}+\zeta^{\varphi_{2}}\frac{\partial}{\partial\varphi_{2}},
\]
where all coefficients depend on $(t,\xi,\eta,\varphi_{1},\varphi_{2})$.
Notice that the function $S(\xi,\eta)$ is considered to be given.
Calculations show that the admitted generator has the form
\[
X=\sum_{j=1}^{10}c_{i}X_{j}+X_{h},
\]
and the classifying equation is
\begin{equation}
h_{\xi}S_{\eta}-(h_{\eta}-2\gamma c_{9}\xi)S_{\xi}=2\gamma c_{10}S,\label{eq:classifying}
\end{equation}
where $c_{7}(\gamma-2)=0$, and
\[
X_{1}=\frac{\partial}{\partial\varphi_{1}},\,\,\,\,X_{2}=\frac{\partial}{\partial\varphi_{2}},\,\,\,X_{3}=t\frac{\partial}{\partial\varphi_{1}},\,\,\,X_{4}=t\frac{\partial}{\partial\varphi_{2}},
\]
\[
X_{5}=\varphi_{2}\frac{\partial}{\partial\varphi_{1}}-\varphi_{1}\frac{\partial}{\partial\varphi_{2}},\,\,\,X_{6}=\frac{\partial}{\partial t},
\]
\[
X_{7}=t\left(t\frac{\partial}{\partial t}+\varphi_{1}\frac{\partial}{\partial\varphi_{1}}+\varphi_{2}\frac{\partial}{\partial\varphi_{2}}\right),\,\,\,X_{8}=\gamma t\frac{\partial}{\partial t}+\varphi_{1}\frac{\partial}{\partial\varphi_{1}}+\varphi_{2}\frac{\partial}{\partial\varphi_{2}},
\]
\[
X_{9}=2\gamma\xi\frac{\partial}{\partial\xi}+(\gamma-1)(\varphi_{1}\frac{\partial}{\partial\varphi_{1}}+\varphi_{2}\frac{\partial}{\partial\varphi_{2}}),\,\,\,X_{10}=\varphi_{1}\frac{\partial}{\partial\varphi_{1}}+\varphi_{2}\frac{\partial}{\partial\varphi_{2}},
\]
\[
X_{h}=-h_{\eta}\frac{\partial}{\partial\xi}+h_{\xi}\frac{\partial}{\partial\eta}.
\]
The analysis of the classifying equation is split into two cases: isentropic
and nonisentropic flows.

\section{Isentropic flows}

In this case $S_{\xi}=0$, $S_{\eta}=0$, and one derives from the
classifying equation that
\[
c_{10}=0.
\]
Thus, for general choice of $\gamma$, the admitted Lie algebra is
infinite dimensional: $L_{8}\oplus\{X_{h}\}$, where a basis of $L_{8}$
consists of the generators $X_{j}$ $(j=1,2,...,9;\,\,\,j\neq7)$,
and $h(\xi,\eta)$ is an arbitrary function. For $\gamma=2$ the basis
of the finite part $L_{8}$ is extended by the generator $X_{7}$.

\subsection{Conservation laws in Lagrangian coordinates}

For a symmetry to be divergent (variational) one needs to require
that
\begin{equation}
c_{9}=\frac{(\gamma-2)}{2(2\gamma-1)}c_{8}.\label{eq:Nov22.1}
\end{equation}

For a divergent symmetry one obtains

\[
b_{1}=c_{4}\varphi_{2}+\frac{1}{\gamma(\gamma-1)}c_{7}(\varphi_{1}^{2}+\varphi_{2}^{2})+c_{3}\varphi_{1}\ \ b_{2}=0,\ \ b_{3}=0.
\]
This gives that each of the generators $X_{j}$, $(j=1,2,3,...,7)$,
$X_{h}$ and
\[
\tilde{X}_{8}=X_{8}+\frac{(\gamma-2)}{2(2\gamma-1)}X_{9}
\]
provides a conservation law. These conservation laws have the conserved
vector coordinates:
\[
T_{1}^{t}=\varphi_{1t},\ \ T_{1}^{\xi}=S\varphi_{2\eta}J^{-\gamma},\ \ T_{1}^{\eta}=-S\varphi_{2\xi}J^{-\gamma};
\]
\[
T_{2}^{t}=\varphi_{2t},\ \ T_{2}^{\xi}=-S\varphi_{1\eta}J^{-\gamma},\ \ T_{2}^{\eta}=S\varphi_{1\xi}J^{-\gamma};
\]
\[
T_{3}^{t}=\varphi_{1t}t-\varphi_{1},\ \ T_{3}^{\xi}=tS\varphi_{2\eta}J^{-\gamma},\ \ T_{3}^{\eta}=-tS\varphi_{2\xi}J^{-\gamma};
\]
\[
T_{4}^{t}=\varphi_{2t}t-\varphi_{2},\ \ T_{4}^{\xi}=-tS\varphi_{1\eta}J^{-\gamma},\ \ T_{4}^{\eta}=tS\varphi_{1\xi}J^{-\gamma};
\]
\[
T_{5}^{t}=\varphi_{1t}\varphi_{2}-\varphi_{1}\varphi_{2t},\ \ T_{5}^{\xi}=(\varphi_{1}\varphi_{1\eta}+\varphi_{2}\varphi_{2\eta})SJ^{-\gamma},\ \ T_{5}^{\eta}=-(\varphi_{1}\varphi_{1\xi}+\varphi_{2}\varphi_{2\xi})S;
\]
\[
\begin{array}{c}
T_{6}^{t}=\varphi_{1t}^{2}+\varphi_{2t}^{2}+\frac{2}{\gamma-1}SJ^{1-\gamma},\ \ T_{6}^{\xi}=2(\varphi_{1t}\varphi_{2\eta}-\varphi_{1\eta}\varphi_{2t})SJ^{-\gamma},\\[2ex]
T_{6}^{\eta}=2(\varphi_{1\xi}\varphi_{2t}-\varphi_{1t}\varphi_{2\xi})SJ^{-\gamma};
\end{array}
\]
%\[
%\begin{array}{c}
%T_{7}^{t}=-\gamma^{-1}\left((\varphi_{2}-\varphi_{2t}t)^{2}+(\varphi_{1}-\varphi_{1t}t)^{2}+2t^{2}SJ^{1-\gamma}\right),\\[2ex]
%T_{7}^{\xi}=J^{-\gamma}((\varphi_{1}-\varphi_{1t}t)\varphi_{2\eta}-(\varphi_{2}-\varphi_{2t}t)\varphi_{1\eta})St),\\[2ex]
%T_{7}^{\eta}=-J^{-\gamma}((\varphi_{1}-\varphi_{1t}t)\varphi_{2\xi}-(\varphi_{2}-\varphi_{2t}t)\varphi_{1\xi})St),
%\end{array}
%\]
\[
\begin{array}{c}
\tilde{T}_{8}^{t}=2(1-\gamma)(\gamma-2)\xi(\varphi_{1t}\varphi_{1\xi}+\varphi_{2t}\xi\varphi_{2\xi})+2(1-2\gamma)tJ^{1-\gamma}S\\
-t(2\gamma-1)(\gamma-1)(\varphi_{1t}^{2}+\varphi_{2t}^{2})+(\gamma^{2}-1)(\varphi_{1t}\varphi_{1}+\varphi_{2t}\varphi_{2}),\\[2ex]
\tilde{T}_{8}^{\xi}=(\gamma-2)\xi((\gamma-1)(\varphi_{1t}^{2}+\varphi_{2t}^{2})-2\gamma J^{1-\gamma}S)\\
+SJ^{-\gamma}(\gamma-1)\left(2(2\gamma-1)t(\varphi_{1\eta}\varphi_{2t}-\varphi_{1t}\varphi_{2\eta})+(\gamma+1)(\varphi_{1}\varphi_{2\eta}-\varphi_{1\eta}\varphi_{2})\right),\\[2ex]
\tilde{T}_{8}^{\eta}=SJ^{-\gamma}(\gamma-1)\left(2(2\gamma-1)t(\varphi_{1t}\varphi_{2\xi}-\varphi_{1\xi}\varphi_{2t})+(\gamma+1)(\varphi_{1\xi}\varphi_{2}-\varphi_{1}\varphi_{2\xi})\right);
\end{array}
\]

\[
T_{h}^{t}=2(\gamma-1)\left((\varphi_{1t}\varphi_{1\xi}+\varphi_{2t}\varphi_{2\xi})h_{\eta}-(\varphi_{1t}\varphi_{1\eta}+\varphi_{2t}\varphi_{2\eta})h_{\xi}\right),
\]
\[
T_{h}^{\xi}=-h_{\eta}((\gamma-1)(\varphi_{1t}^{2}+\varphi_{2t}^{2})-2\gamma SJ^{1-\gamma}),
\]
\[
T_{h}^{\eta}=h_{\xi}((\gamma-1)(\varphi_{1t}^{2}+\varphi_{2t}^{2})-2\gamma J^{1-\gamma}S).
\]
Here and further on the subscripts relate to the corresponding generator
providing the conservation law.

In the case where $\gamma=2$ there is one more conservation law in
Lagrangian coordinates which corresponds to the generator $X_{7}$:
\[
\begin{array}{c}
T_{7}^{t}=((\varphi_{2}-\varphi_{2t}t)^{2}+(\varphi_{1}-\varphi_{1t}t)^{2}+2SJ^{-1}t^{2}),\,\,\,T_{7}^{\xi}=2tSJ^{-2}((\varphi_{2}-\varphi_{2t}t)\varphi_{1\eta}-(\varphi_{1}-\varphi_{1t}t)\varphi_{2\eta}),\\[2ex]
T_{7}^{\eta}=2tSJ^{-2}((\varphi_{1}-\varphi_{1t}t)\varphi_{2\xi}-(\varphi_{2}-\varphi_{2t}t)\varphi_{1\xi}).
\end{array}
\]

\subsection{Conservation laws in Eulerian coordinates}

%%%%%%%%%%%%%%%%%%%%%%%%%%%%%%%%%%%%%%%%%%%%%%%%%%%%%%%%%%%%%%%%%%%%%%%%%%%%%%%
%%%%%%%%%%             operatorS of total derivativeS in Eulerian  %%%%%%%%%%%%
%%%%%%%%%%%%%%%%%%%%%%%%%%%%%%%%%%%%%%%%%%%%%%%%%%%%%%%%%%%%%%%%%%%%%%%%%%%%%%%

First of all, we should mention the conservation law of mass which
is valid for isentropic as well as nonisentropic flows,
\[
\rho_{t}+(\rho u)_{x}+(\rho v)_{y}=0.
\]
The latter conserved vectors $(\tilde{T}_{i}^{t}$,$\tilde{T}_{i}^{\xi},\tilde{T}_{i}^{\eta})$,
$(i=1,2,...,8)$ have the following forms in Eulerian coordinates
\[
^{e}T_{1}^{t}=\rho u,\ \ T_{1}^{x}=\rho u^{2}+\rho^{\gamma}S,\ \ T_{1}^{y}=\rho uv;
\]
\[
^{e}T_{2}^{t}=\rho v,\ \ T_{2}^{x}=\rho uv,\ \ T_{2}^{y}=\rho v^{2}+\rho^{\gamma}S;
\]
\[
^{e}T_{3}^{t}=\rho(tu-x),\ \ T_{3}^{x}=\rho u(tu-x)+t\rho^{\gamma}S,\ \ T_{3}^{y}=\rho v(tu-x);
\]
\[
^{e}T_{4}^{t}=\rho(tv-y),\ \ T_{4}^{x}=\rho u(tv-y),\ \ T_{4}^{y}=\rho v(tv-y)+t\rho^{\gamma}S;
\]
\[
^{e}T_{5}^{t}=\rho(uy-vx),\ \ T_{5}^{x}=\rho u(uy-vx)+y\rho^{\gamma}S,\ \ T_{5}^{y}=\rho v(uy-vx)-x\rho^{\gamma}S;
\]
\[
\begin{array}{c}
^{e}T_{6}^{t}=\rho\frac{u^{2}+v^{2}}{2}+\frac{1}{\gamma-1}\rho^{\gamma}S,\,\,\,T_{6}^{x}=u\left(\rho\frac{u^{2}+v^{2}}{2}+\frac{1}{\gamma-1}\rho^{\gamma}S\right),\\[2ex]
T_{6}^{y}=v\left(\rho\frac{u^{2}+v^{2}}{2}+\frac{1}{\gamma-1}\rho^{\gamma}S\right).
\end{array}
\]
The function\footnote{According to a previous remark, the sign $\tilde{}$ is omitted.}
$h(\xi,\eta)$ becomes a function $h(t,x,y)$ which only satisfies
the equation
\begin{equation}
h_{t}+uh_{x}+vh_{y}=0.\label{eq:ad_fun}
\end{equation}
The corresponding conservation law is
\begin{equation}
\begin{array}{c}
^{e}T_{h}^{t}=2(\gamma-1)(uh_{y}-vh_{x}),\ \ T_{h}^{x}=(\gamma-1)((u^{2}-v^{2})h_{y}-2uvh_{x})+2\gamma\rho^{\gamma-1}h_{y}S,\\
T_{h}^{y}=(\gamma-1)((u^{2}-v^{2})h_{x}+2uvh_{y})-2\gamma\rho^{\gamma-1}h_{x}S.
\end{array}\label{eq:ad_cons_law_h}
\end{equation}
The conserved vector $(\tilde{T}_{8}^{t}$,$\tilde{T}_{8}^{\xi},\tilde{T}_{8}^{\eta})$
cannot be represented in Eulerian coordinates because of the presence
of $\xi$. Also notice that for $\gamma=2$ the term containing $\xi$
vanishes.

For $\gamma=2$ there are two more conserved vectors: the conserved
vector $(T_{7}^{t},T_{7}^{\xi},T_{7}^{\eta})$ becomes
\[
\begin{array}{c}
^{e}T_{7}^{t}=\rho\left(2t(ux+vy)-(x^{2}+y^{2})-t^{2}(u^{2}+v^{2}+2\rho S)\right),\\
T_{7}^{x}=\rho\left(u(2t(ux+vy)-(x^{2}+y^{2})-(u^{2}+v^{2})t^{2})-2t\rho S(2tu-x)\right),\\
T_{7}^{y}=\rho\left(v(2t(ux+vy)-(x^{2}+y^{2})-(u^{2}+v^{2})t^{2})-2t\rho S(2tv-y)\right),
\end{array}
\]
and, as noted above, for $\gamma=2$ the term with $\xi$ vanishes
in the conserved vector $(\tilde{T}_{8}^{t}$,$\tilde{T}_{8}^{\xi},\tilde{T}_{8}^{\eta})$,
and the conserved vector $(\tilde{T}_{8}^{t}$,$\tilde{T}_{8}^{\xi},\tilde{T}_{8}^{\eta})$
also has its representation in Eulerian coordinates:
\[
\begin{array}{c}
^{e}T_{8}^{t}=\rho(t(u^{2}+v^{2})-(ux+vy)+2t\rho S),\,\,\,T_{8}^{x}=\rho(u(t(u^{2}+v^{2})-(ux+vy))+(4tu-x)\rho S),\\
T_{8}^{y}=\rho(v(t(u^{2}+v^{2})-(ux+vy))+(4tv-y)\rho S).
\end{array}
\]

\subsection{Discussion}

First of all we note that for deriving relation (\ref{eq:Nov22.1})
it was important to consider generator $X$ in its general form: not
analyzing condition (\ref{eq:divergent}) for each specific generator.

The conserved vectors $(T_{i}^{t},T_{i}^{x},T_{i}^{y})$, $(i=1,2,...,8)$
are well-known \citep{bk:Ibragimov[1983]}: the vectors $(T_{1}^{t},T_{1}^{x},T_{1}^{y})$
and $(T_{2}^{t},T_{2}^{x},T_{2}^{y})$ give conservation laws of linear
momentum; the vectors $(T_{3}^{t},T_{3}^{x},T_{3}^{y})$ and $(T_{4}^{t},T_{4}^{x},T_{4}^{y})$
correspond to the conservation law of angular momentum motion
of the center of mass; the vector $(T_{6}^{t},T_{6}^{x},T_{6}^{y})$ corresponds
to the conservation law of energy; the vectors $(T_{7}^{t},T_{7}^{x},T_{7}^{y})$
and $(T_{8}^{t},T_{8}^{x},T_{8}^{y})$ are extensions of the classical
conservation laws for a polytropic gas with $\gamma=2$ derived in
\citep{bk:Ibragimov[1983]}.

The conserved vector $(T_{h}^{t},T_{h}^{x},T_{h}^{y})$ with the Lagrangian
invariant $h(t,x,y)$ (satisfying equation (\ref{eq:ad_fun})) provides
a new conservation law. In contrast to \citep{bk:SjobergMahomed2004}
this conservation law is local. It should be also noted that this
conservation law is naturally derived: its counterpart in Lagrangian
coordinates was derived directly using Noether's theorem without any
additional constructions.

{\bf Remark}.
The gas dynamics equations (\ref{eq:cons_mass_E})-(\ref{eq:entropy_E}) for an isentropic flow and $\gamma=2$,  coincide with the hyperbolic shallow water equations. Group properties of the shallow water equations describing flows over a bed which is rotating
with position-dependent Coriolis parameter in Lagrangian coordinates\footnote{If the Coriolis parameter vanishes, then the equations considered in \citep{bk:BilaMansfieldClarkson2006} coincide with equations (\ref{eq:cons_mass_L})-(\ref{eq:geometrical}).}
are studied in \citep{bk:BilaMansfieldClarkson2006}. Conservation laws were constructed by using a Lagrangian of the form presented in \citep{bk:Salmon1983}. In the present paper the admitted Lie group is wider, and we also obtained more conservation laws. This can be explained by the presence of nonzero Coriolis parameter and different Lagrangian used in \citep{bk:BilaMansfieldClarkson2006}.

\section{Nonisentropic flows}

\subsection{Admitted Lie algebra}

Choosing the functions $\psi_{1}(\xi,\eta)$, $\psi_{2}(\xi,\eta)$
and $\psi_{0}(\xi,\eta)$ satisfying the equations
\[
\psi_{1\eta}S_{\xi}-(\psi_{1\xi}S_{\eta}+\xi S_{\xi})=0,
\]
\[
S_{\xi}\psi_{2\eta}-S_{\eta}\psi_{2\xi}+2S=0,
\]
\begin{equation}
\psi_{0\eta}S_{\xi}-\psi_{0\xi}S_{\eta}=0,\label{eq:Nov17.1}
\end{equation}
one finds the general solution of the classifying equation (\ref{eq:classifying}),
\[
h=\psi_{0}+\gamma(2c_{9}\psi_{1}+c_{10}\psi_{2}).
\]
Notice that the general solution of (\ref{eq:Nov17.1}) is $\psi_{0}=F(S)$,
where the function $F$ is an arbitrary function.

It is convenient to introduce
\[
c_{10}=\frac{\gamma-2}{2}c_{8}+\tilde{c}_{10}.
\]
Then the admitted generator of equations (\ref{eq:gas_L_1}), (\ref{eq:gas_L_2})
has the form
\[
X=\sum_{j=1}^{7}c_{i}X_{j}+c_{8}\tilde{X}_{8}+c_{9}\tilde{X}_{9}+\tilde{c}_{10}\tilde{X}_{10}+X_{\psi_{0}},
\]
where $c_{7}(\gamma-2)=0$, and
\[
X_{1}=\frac{\partial}{\partial\varphi_{1}},\,\,\,\,X_{2}=\frac{\partial}{\partial\varphi_{2}},\,\,\,X_{3}=t\frac{\partial}{\partial\varphi_{1}},\,\,\,X_{4}=t\frac{\partial}{\partial\varphi_{2}},
\]
\[
X_{5}=\varphi_{2}\frac{\partial}{\partial\varphi_{1}}-\varphi_{1}\frac{\partial}{\partial\varphi_{2}},\,\,\,X_{6}=\frac{\partial}{\partial t},\,\,\,X_{7}=t\left(t\frac{\partial}{\partial t}+\varphi_{1}\frac{\partial}{\partial\varphi_{1}}+\varphi_{2}\frac{\partial}{\partial\varphi_{2}}\right),
\]
\[
\tilde{X}_{8}=\frac{\gamma}{2}\left(2t\frac{\partial}{\partial t}-(\gamma-2)\left(\psi_{2\eta}\frac{\partial}{\partial\xi}-\psi_{2\xi}\frac{\partial}{\partial\eta}\right)+\varphi_{1}\frac{\partial}{\partial\varphi_{1}}+\varphi_{2}\frac{\partial}{\partial\varphi_{2}}\right),
\]
\[
\tilde{X}_{9}=2\gamma\left((\xi-\psi_{1\eta})\frac{\partial}{\partial\xi}+\psi_{1\xi}\frac{\partial}{\partial\eta}\right)+(\gamma-1)\left(\varphi_{1}\frac{\partial}{\partial\varphi_{1}}+\varphi_{2}\frac{\partial}{\partial\varphi_{2}}\right),
\]
\[
\tilde{X}_{10}=\gamma\left(-\psi_{2\eta}\frac{\partial}{\partial\xi}+\psi_{2\xi}\frac{\partial}{\partial\eta}\right)+\varphi_{1}\frac{\partial}{\partial\varphi_{1}}+\varphi_{2}\frac{\partial}{\partial\varphi_{2}},
\]
\[
X_{F}=F^{\prime}\left(S_{\xi}\frac{\partial}{\partial\eta}-S_{\eta}\frac{\partial}{\partial\xi}\right).
\]

\subsection{Conservation laws in mass Lagrangian coordinates }

The condition for the generator $X$ to be divergent (variational)
gives that
\[
c_{9}=-\frac{1}{2\gamma-1}\tilde{c}_{10}
\]
and
\[
b_{1}=c_{3}\varphi_{1}+c_{4}\varphi_{2}+\frac{c_{7}}{2(\gamma-1)}(\varphi_{1}^{2}
+\varphi_{2}^{2}),\,\,\,b_{2}=0,\ \ b_{3}=0,
\]
\[
\begin{array}{c}
\hat{X}_{9}=\tilde{X}_{10}-\frac{1}{2\gamma-1}\tilde{X}_{9}
\\
\displaystyle
=\frac{\gamma}{2\gamma-1}\left(\left(2(\psi_{1\eta}-\xi)-(2\gamma-1)\psi_{2\eta}\right)\frac{\partial}{\partial\xi}+\left(2\psi_{1\xi}+(2\gamma-1)\psi_{2\xi}\right)\frac{\partial}{\partial\eta}+\varphi_{1}\frac{\partial}{\partial\varphi_{1}}+\varphi_{2}\frac{\partial}{\partial\varphi_{2}}\right).
\end{array}
\]
The conserved vectors for $(T_{i}^{t},T_{i}^{\xi},T_{i}^{\eta})$,
$(i=1,2,...,6)$ are the same as in the isentropic case. The remaining
conserved vectors are

\[
T_{8}^{t}=(\gamma-2)\left((\varphi_{1t}\varphi_{1\xi}+\varphi_{2t}\varphi_{2\xi})\psi_{2\eta}-(\varphi_{1t}\varphi_{1\eta}+\varphi_{2t}\varphi_{2\eta})\psi_{2\xi}\right)
\]
\[
+(\varphi_{1}-\varphi_{1t}t)\varphi_{1t}+(\varphi_{2}-\varphi_{2t}t)\varphi_{2t}-\frac{2}{\gamma-1}tJ^{1-\gamma}S,
\]
\[
T_{8}^{\xi}=\left((\varphi_{1}-2\varphi_{1t}t)\varphi_{2\eta}-(\varphi_{2}-2\varphi_{2t}t)\varphi_{1\eta}\right)SJ^{-\gamma}-\frac{\gamma-2}{2}\psi_{2\eta}\left(\varphi_{1t}^{2}+\varphi_{2t}^{2}-\frac{2\gamma}{\gamma-1}J^{1-\gamma}S\right),
\]
\[
T_{8}^{\eta}=-\left((\varphi_{1}-2\varphi_{1t}t)\varphi_{2\xi}-(\varphi_{2}-2\varphi_{2t}t)\varphi_{1\xi}\right)SJ^{-\gamma}+\frac{\gamma-2}{2}\psi_{2\xi}\left(\varphi_{1t}^{2}+\varphi_{2t}^{2}-\frac{2\gamma}{\gamma-1}J^{1-\gamma}S\right);
\]

\[
\begin{array}{c}
T_{9}^{t}=-((2\gamma-1)\psi_{2\xi}-2\psi_{1\xi})(\varphi_{1t}\varphi_{1\eta}+\varphi_{2t}\varphi_{2\eta})\\
+((2\gamma-1)\psi_{2\eta}+2\xi-2\psi_{1\eta})(\varphi_{1t}\varphi_{1\xi}+\varphi_{2t}\varphi_{2\xi})+(\varphi_{1}\varphi_{1t}+\varphi_{2}\varphi_{2t})),
\end{array}
\]
\[
T_{9}^{\xi}=\frac{1}{2}\left((2\psi_{1\eta}-2\xi-(2\gamma-1)\psi_{2\eta})(\varphi_{1t}^{2}+\varphi_{2t}^{2}-\frac{2\gamma}{\gamma-1}J^{1-\gamma}S)+2J^{-\gamma}S(\varphi_{1}\varphi_{2\eta}-\varphi_{1\eta}\varphi_{2})\right),
\]
\[
T_{9}^{\eta}=\frac{1}{2}\left(((2\gamma-1)\psi_{2\xi}-2\psi_{1\xi})(\varphi_{1t}^{2}+\varphi_{2t}^{2}-\frac{2\gamma}{\gamma-1}J^{1-\gamma}S)-2J^{-\gamma}S(\varphi_{1}\varphi_{2\xi}-\varphi_{1\xi}\varphi_{2})\right);
\]
\[
T_{F}^{t}=F^{\prime}\left((\varphi_{1t}\varphi_{1\xi}+\varphi_{2t}\varphi_{2\xi})S_{\eta}-(\varphi_{1t}\varphi_{1\eta}+\varphi_{2t}\varphi_{2\eta})S_{\xi}\right),
\]
\[
T_{F}^{\xi}=-\frac{1}{2}F^{\prime}S_{\eta}(\varphi_{1t}^{2}+\varphi_{2t}^{2}-\frac{2\gamma}{\gamma-1}J^{1-\gamma}S),\,\,\,T_{\psi_{0}}^{\eta}=\frac{1}{2}F^{\prime}S_{\xi}(\varphi_{1t}^{2}+\varphi_{2t}^{2}-\frac{2\gamma}{\gamma-1}J^{1-\gamma}S).
\]

In the case $\gamma=2$ there is one more conserved vector corresponding
to the generator $X_{7}$. This conserved vector is the vector $(T_{7}^{t},T_{7}^{\xi},T_{7}^{\eta})$
as in the isentropic case for $\gamma=2$:
\[
\begin{array}{c}
T_{7}^{t}=((\varphi_{2}-\varphi_{2t}t)^{2}+(\varphi_{1}-\varphi_{1t}t)^{2}+2SJ^{-1}t^{2}),\\
T_{7}^{\xi}=2tSJ^{-2}((\varphi_{2}-\varphi_{2t}t)\varphi_{1\eta}-(\varphi_{1}-\varphi_{1t}t)\varphi_{2\eta}),\\
T_{7}^{\eta}=2tSJ^{-2}((\varphi_{1}-\varphi_{1t}t)\varphi_{2\xi}-(\varphi_{2}-\varphi_{2t}t)\varphi_{1\xi}).
\end{array}
\]

\subsection{Conservation laws in Eulerian coordinates}

%%%%%%%%%%%%%%%%%%%%%%%%%%%%%%%%%%%%%%%%%%%%%%%%%%%%%%%%%%%%%%%%%%%%%%%%%%%%%%%
%%%%%%%%%%             operatorS of total derivativeS in Eulerian  %%%%%%%%%%%%
%%%%%%%%%%%%%%%%%%%%%%%%%%%%%%%%%%%%%%%%%%%%%%%%%%%%%%%%%%%%%%%%%%%%%%%%%%%%%%%

Representations of the conserved vectors $(T_{i}^{t},T_{i}^{\xi},T_{i}^{\eta})$,
$(i=1,2,...,7)$ are of similar forms $(^{e}T_{i}^{t},T_{i}^{x},T_{i}^{y})$
as in the isentropic case. The other conserved vectors in Eulerian
coordinates have the form
\[
\begin{array}{c}
^{e}T_{8}^{t}=(\gamma-2)\left(\psi_{2y}u-\psi_{2x}v\right)-\left(\rho(t(u^{2}+v^{2})-(ux+vy))+\frac{2}{\gamma-1}t\rho^{\gamma}S\right),\\[2ex]
T_{8}^{x}=-\left(\rho^{\gamma}S(\frac{2\gamma}{\gamma-1}tu-x)+\rho u(t(u^{2}+v^{2})-(ux+vy))\right)\\
+\frac{\gamma-2}{2}\left(\psi_{2y}(u^{2}-v^{2}+\frac{2\gamma}{\gamma-1}\rho^{\gamma-1}S)-2\psi_{2x}uv\right),\\[2ex]
T_{8}^{y}=-\left(\rho^{\gamma}S(\frac{2\gamma}{\gamma-1}tv-y)+\rho v(t(u^{2}+v^{2})-(ux+vy))\right)\\
+\frac{\gamma-2}{2}\left(\psi_{2x}(u^{2}-v^{2}-\frac{2\gamma}{\gamma-1}\rho^{\gamma-1}S)+2\psi_{2y}uv\right);
\end{array}
\]
\[
\begin{array}{c}
^{e}T_{F}^{t}=2F^{\prime}(S_{y}u-S_{x}v),\ \ T_{F}^{x}=F^{\prime}((u^{2}-v^{2})S_{y}-2S_{x}uv+\frac{2\gamma}{\gamma-1}\rho^{\gamma-1}SS_{y}),\\[2ex]
T_{F}^{y}=F^{\prime}((u^{2}-v^{2})S_{x}+2S_{y}uv-\frac{2\gamma}{\gamma-1}\rho^{\gamma-1}SS_{x}).
\end{array}
\]
Here the function $\psi_{2}(t,x,y)$ satisfies the equations
\begin{equation}
\psi_{2t}+u\psi_{2x}+v\psi_{2y}=0,\,\,\,\psi_{2x}S_{y}-\psi_{2y}S_{x}=2\rho S.\label{eq:Nov17.2}
\end{equation}

The conserved vector $(T_{9}^{t},T_{9}^{\xi},T_{9}^{\eta})$ has no
a representation in Eulerian coordinates.

\subsection{Discussion}

Continuing the discussions presented in the previous section, the
conserved vectors $(^{e}T_{8}^{t},T_{8}^{x},T_{8}^{y})$ and  $(^{e}T_{F}^{t},T_{F}^{x},T_{F}^{y})$
provide new conservation laws. The conserved vector $(^{e}T_{F}^{t},T_{F}^{x},T_{F}^{y})$
is similar to the vector $(^{e}T_{h}^{t},T_{h}^{x},T_{h}^{y})$ studied
for isentropic flows.

The conserved vector $(^{e}T_{8}^{t},T_{8}^{x},T_{8}^{y})$ contains
the function $\psi_{2}(t,x,y)$ which satisfies the equations (\ref{eq:Nov17.2}).
The overdetermined system of the gas dynamics equations with (\ref{eq:Nov17.2})
is involutive. This leads to a similar idea as in \citep{bk:SjobergMahomed2004}
to extend an original set of the dependent variables, and to seek
for new conservation laws containing the added variables. However,
in the present cases the conservation laws are naturally derived:
their counterparts in Lagrangian coordinates were derived directly
using Noether's theorem without any additional assumptions.

\section{Conclusion}

New conservation laws of two-dimensional gas dynamics equations of a polytropic gas are found in the present paper. The conservation laws are derived using corresponding Lagrangian and Noether's theorem. For constructing conservation laws we used the complete group classification of the Euler-Lagrange equations (\ref{eq:gas_L_1}), (\ref{eq:gas_L_2}). In contrast to the one-dimensional case of the Lagrangian gas dynamics equations \citep{bk:DorodnitsynKozlovMeleshko2019}, where $\varphi_{1\eta}=0$ and $\varphi_{2}=\eta$, the group classification only separates out the gas dynamics equations (\ref{eq:gas_L_1}), (\ref{eq:gas_L_2}) on analysis of isentropic and nonisentropic flows, and the admitted Lie group found has no constraints on the entropy\footnote{See also \citep{bk:WebbZank[2009]}.}.
The complete set of admitted generators allowed us to use Noether's theorem for deriving conservation laws in mass Lagrangian coordinates. Their corresponding counterparts in Eulerian coordinates were also constructed. Some of these counterparts contain Lagrangian invariants. Moreover, the conserved vector $(^{e}T_{8}^{t},T_{8}^{x},T_{8}^{y})$ contains the function $\psi_2(t,x,y)$ which satisfies an overdetermined system of equations. One can show that the overdetermined system of equations consisting of these equations and the gas dynamics equations is involutive. Applying the group analysis method to this overdetermined system the admitted Lie group of the gas dynamics can be extended.
The involutiveness of this system also gives a similar idea as in \citep{bk:SjobergMahomed2004}:
to extend an original set of the dependent variables, and to seek
for new conservation laws containing the added variables. However,
in the present paper the conservation laws are naturally derived:
their counterparts in Lagrangian coordinates were derived directly
using Noether's theorem without any additional assumptions.

It should be also noted that there are conservation laws which have no their counterpart in Eulerian coordinates.

\section*{Acknowledgements}
The research  was supported by Russian Science Foundation Grant No 18-11-00238
`Hy\-dro\-dynamics-type equations: symmetries, conservation laws, invariant difference schemes'.
E.I.K. also acknowledges Suranaree University of Technology for %support through
Full-time Master Researcher Fellowship (15/2561).
The authors thank V.A.Do\-rod\-ni\-tsyn and E.Schulz for valuable discussions.

%\section*{References}

%\bibliographystyle{unsrt}
%\bibliography{references}

\begin{thebibliography}{10}

\bibitem{bk:Sedov[mss]}
L.~I. Sedov.
\newblock {\em Continuum mechanics, v. 1, 5-th ed.}
\newblock Nauka, Moscow, 1994.
\newblock in Russian.

\bibitem{bk:Webb2018}
G.~Webb.
\newblock {\em Magnetohydrodynamics and Fluid Dynamics: Action Principles and
  Conservation Laws}.
\newblock Springer, Heidelberg, 2018.
\newblock Lecture Notes in Physics, v. 946.

\bibitem{bk:WebbZank[2009]}
G.~M. Webb and G.~P. Zank.
\newblock Scaling symmetries, conservation laws and action principles in
  one-dimensional gas dynamics.
\newblock {\em Journal of Physics A: Mathematical and Theoretical}, 42, 2009.
\newblock Paper 475205.

\bibitem{bk:DespresMazeran2005}
B.~Despr\'es and C.~Mazeran.
\newblock {L}agrangian gas dynamics in two dimensions and {L}agrangian systems.
\newblock {\em Arch. Rational Mech. Anal.}, 178:327--372, 2005.

\bibitem{bk:RozhdYanenko[1978]}
B.~L. Rozhdestvenskii and N.~N. Yanenko.
\newblock {\em Systems of quasilinear equations and their applications to gas
  dynamics, 2nd ed.}
\newblock Nauka, Moscow, 1978.
\newblock {E}nglish translation published by Amer. Math. Soc., Providence, RI,
  1983.

\bibitem{bk:SiriwatKaewmaneeMeleshko2016}
P.~Siriwat, C.~Kaewmanee, and S.~V. Meleshko.
\newblock Symmetries of the hyperbolic shallow water equations and the
  {G}reen-{N}aghdi model in {L}agrangian coordinates.
\newblock {\em International Journal of Non-Linear Mechanics}, 86:185--195,
  2016.

\bibitem{bk:VorakaKaewmaneeMeleshko2019}
P.~Voraka, C.~Kaewmanee, and S.~V. Meleshko.
\newblock Symmetries of the shallow water equations in the {B}oussinesq
  approximation.
\newblock {\em Commun Nonlinear Sci Numer Simulat}, 67:1--12, 2019.

\bibitem{bk:Ovsiannikov1978}
L.~V. Ovsiannikov.
\newblock {\em Group analysis of differential equations}.
\newblock Nauka, Moscow, 1978.
\newblock {E}nglish translation, {A}mes, {W}.{F}., Ed., published by Academic
  Press, New York, 1982.

\bibitem{bk:Olver[1986]}
P.~J. Olver.
\newblock {\em Applications of {L}ie groups to differential equations}.
\newblock Springer-Verlag, New York, 1986.

\bibitem{bk:MarsdenRatiu[1994]}
J.~Marsden and T.~Ratiu.
\newblock {\em Introduction to Mechanics and Symmetry}.
\newblock Springer-Verlag, New York, 1994.

\bibitem{bk:Ibragimov[1999]}
N.~H. Ibragimov.
\newblock {\em Elementary {L}ie Group Analysis and Ordinary Differential
  Equations}.
\newblock Wiley \& Sons, Chichester, 1999.

\bibitem{bk:Cantwell[2002]}
B.~J. Cantwell.
\newblock {\em Introduction to symmetry analysis}.
\newblock Cambridge University Press, Cambridge, 2002.

\bibitem{bk:Ovsiannikov[1994]}
L.~V. Ovsiannikov.
\newblock Program {SUBMODELS}. {G}as dynamics.
\newblock {\em J. Appl. Maths Mechs}, 58(4):30--55, 1994.

\bibitem{bk:AkhatovGazizovIbragimov[1991]}
I.~S. Akhatov, R.~K. Gazizov, and N.~H. Ibragimov.
\newblock Nonlocal symmetries. {H}euristic approach.
\newblock {\em J. Math. Sci.}, 55(1):1401--1450, 1991.
\newblock Journal of Soviet Mathematics (in Russian).

\bibitem{bk:HandbookLie_v2}
N.~H. Ibragimov, editor.
\newblock {\em {CRC} Handbook of {L}ie Group Analysis of Differential
  Equations}, volume~2.
\newblock CRC Press, Boca Raton, 1995.

\bibitem{bk:SjobergMahomed2004}
A.~Sj\"oberg and F.~M. Mahomed.
\newblock Non-local symmetries and conservation laws for one-dimensional gas
  dynamics equations.
\newblock {\em Applied Mathematics and Computation}, 150:379–397, 2004.

\bibitem{bk:DorodnitsynKozlovMeleshko2019}
V.~A. Dorodnitsyn, R.~Kozlov, and S.~V. Meleshko.
\newblock Analysis of 1{D} gas dynamics equations of a polytropic gas in
  {L}agrangian coordinates: symmetry classification, conservation laws,
  difference schemes.
\newblock {\em Commun Nonlinear Sci Numer Simulat}.
\newblock in press.

\bibitem{bk:BilaMansfieldClarkson2006}
N.~Bila, E.~L. Mansfield, and P.~A. Clarkson.
\newblock Symmetry group analysis of the shallow water and semi-geostrophic
  equations.
\newblock {\em The Quarterly Journal of Mechanics and Applied Mathematics},
  59(1):95--123, 2006.

\bibitem{bk:Salmon1983}
R.~Salmon.
\newblock Practical use of hamiltonian's principle.
\newblock {\em J. Fluid Mech.}, 132:431--444, 1983.

\bibitem{bk:Noether[1918]}
E.~Noether.
\newblock Invariante {V}ariationsprobleme.
\newblock {\em Nachr. d. {K}\"oniglichen Gesellschaft der Wissenschaften zu
  G\"ottingen, Math-phys. Klasse}, pages 235--257, 1918.
\newblock English translation in: Transport Theory and Statistical Physics,
  vol. 1, No. 3, 1971, 186-207 (arXiv:physics/0503066 [physics.hist-ph]).

\bibitem{bk:Shmyglevski}
Yu.~D. Shmyglevski.
\newblock {\em Analytical study of gas dynamics and fluid}.
\newblock Editorial URSS, Moscow, 1999.
\newblock in Russian.

\bibitem{Ibragimov2[2007]}
N.~H. Ibragimov.
\newblock A new conservation theorem.
\newblock {\em J. Math. Anal. Appl.}, 333:311--328, 2007.

\bibitem{bk:BlumanCheviakovAnco}
G.~W. Bluman, A.~F. Cheviakov, and S.~C. Anco.
\newblock {\em Applications of Symmetry Methods to Partial Differential
  Equations}.
\newblock Springer, New York, 2010.
\newblock Applied Mathematical Sciences, Vol.168.

\bibitem{bk:SeligerWhitham[1968]}
R.~L. Seliger and G.~B. Whitham.
\newblock Variational principles in continuum mechanics.
\newblock {\em Proc. R. Soc. London. A}, 305:1--25, 1968.

\bibitem{bk:Ibragimov[2010]}
N.~H. Ibragimov.
\newblock Nonlinear self-adjointness in constructing conservation laws.
\newblock {\em Archives of ALGA}, 7/8:1--99, 2010-2011.

\bibitem{bk:DorodnitsynKozlov[2011]}
V.~A. Dorodnitsyn and R.~Kozlov.
\newblock {L}agrangian and {H}amiltonian formalism for discrete equations:
  Symmetries and first integrals.
\newblock In {\em Symmetries and Integrability of Difference Equations}, pages
  7--49. Cambridge University Press, Cambridge, 2011.
\newblock London Mathematical Society Lecture Notes.

\bibitem{bk:Ovsiannikov[2003]}
L.~V. Ovsiannikov.
\newblock {\em Lectures on the gas dynamics equations}.
\newblock Institute of computer studies, Moscow, Izhevsk, 2003.
\newblock Second edition.

\bibitem{bk:Ibragimov[1983]}
N.~H. Ibragimov.
\newblock {\em Transformation Groups Applied to Mathematical Physics}.
\newblock Nauka, Moscow, 1983.
\newblock {E}nglish translation, Reidel, D., Ed., Dordrecht, 1985.

\end{thebibliography}

\end{document}